\documentclass[cpp,a4paper,fleqn%
]{w-art}
\usepackage{times,cite,w-thm}
\usepackage{amsmath}
\usepackage{bm}
\usepackage[]{graphicx}

\begin{document}
\DOIsuffix{theDOIsuffix}
\Volume{46}
\Month{01}
\Year{2007}
\pagespan{1}{}
\Receiveddate{XXXX}
\Reviseddate{XXXX}
\Accepteddate{XXXX}
\Dateposted{XXXX}



\title[Influence of Laser Pulse Parameters on the Properties of $e^-e^+$ Plasmas]
{Influence of Laser Pulse Parameters on the Properties of $\mathbf{e}^- \mathbf{e}^+ $  Plasmas Created from Vacuum}

\author[D.~B.~Blaschke et al]{D.~B.~Blaschke\inst{1,2,}%
  \footnote{Corresponding author\quad E-mail:~\textsf{blaschke@ift.uni.wroc.pl}}}
\address[\inst{1}]{Institute for Theoretical Physics, University of Wroc{\l}aw,
50-204 Wroc{\l}aw, Poland}
\address[\inst{2}]{Bogoliubov Laboratory for Theoretical Physics, Joint Institute for Nuclear Research, RU - 141980 Dubna, Russia}
%
\author[]{B.~K\"ampfer\inst{3,4}}
\address[\inst{3}]{Helmholtzzentrum Dresden-Rossendorf, PF 510119, 01314 Dresden, Germany}
\address[\inst{4}]{Institut f\"ur Theoretische Physik, TU Dresden, 01062 Dresden, Germany}
\author[]{A.~D.~Panferov\inst{5}}
\address[\inst{5}]{Saratov State University, RU - 410026
Saratov, Russia}
\author[]{A.~V.~Prozorkevich\inst{5}}
\author[]{S.~A.~Smolyansky\inst{5,}\footnote{E-mail:~\textsf{smol@sgu.ru}}}

\begin{abstract}
We use the low density approximation within the kinetic theory approach to
vacuum creation of an $e^-e^+$ pair plasma (EPPP) in a strong laser field
in order to investigate the dependence of the
observed EPPP on the form and parameters of a single laser pulse.
The EPPP distribution function is calculated for an arbitrary time dependence
of the electric field in the multiphoton domain (adiabaticity parameter
$\gamma \gg 1$).
The dependence on the field strength, the form and spectrum of the field
pulse is investigated on the basis both analytical and numerical methods.
The obtained results can be useful for examining some observable secondary
processes associated with the dynamical Schwinger effect.
\end{abstract}
\maketitle                   

\section{Introduction \label{sect:in}}

The short wavelength domain of the laser radiation is very prospective for
experimental observation and investigation of the dynamical Schwinger effect
\cite{Blaschke:2005hs,Blaschke:2008wf,Gregori:2010uf}.
The operating $X$-ray laser facilities \cite{Ringwald:2001ib,desy} are able to
generate only very small electric fields in the focal spot with the amplitude
$E_0 \sim 10^{-3} - 10^{-2}~E_c$, where $E_c=m^2/e$ is the Schwinger critical
field strength.
However, there are perspectives to achieve the Schwinger limit in this domain
in the nearest future (e.g., \cite{Bulanov:2003zz,Rhodes}).

In the present work we investigate the residual electron-positron pair plasma
(EPPP) which is generated from vacuum after the cessation of a high-power laser
pulse in the focal spot.
In contrast to the well-known works \cite{Brezin:1970xf,Popov:2001}, our
approach is based on the analysis of a kinetic equation (KE)
\cite{Schmidt:1998vi,Schmidt:1998zh} that is an exact nonperturbative
consequence of the basic equations of motion of QED for the case of a linearly
polarized time dependent electric field ("laser field").
We limit ourselves to the domain of subcritical fields $E_0 \ll E_c$ and to the
case of rather short wavelength $\lambda$ when the adiabacity parameter
$\gamma = 2\pi E_c \lambda_c/E_0\lambda$ is large, $\gamma \gg 1$.
This corresponds to the multiphoton mechanism of the EPPP excitation
($\lambda_c =1/m$ is the Compton wavelength).

In problems of vacuum particle production, one usually studies the
characteristics of the out-state, when the external field is switched off.
This is due to the needs of an experiment aimed at finding free real particles.
Recently, however, there is a growing interest in the properties of the
intermediate states (mid-states), characterized by quasiparticle interactions
in the presence of an external field.
This is motivated by the likely role of the elementary reactions of the
quasiparticles in the field and their possible experimental manifestations
\cite{Blaschke:2005hs,Blaschke:2011is,Bell:2008zzb,Fedotov:2010ja}.

The most complete description of the pair production process for both the mid-
and the out-states is given by the momentum distribution function.
Depending on the shape of the external field the vacuum excitations can form
very complex and diverse momentum distributions which are essentially
nonequilibrium ones.
The complex form of the pair momentum spectrum should significantly affect the
cross sections of elementary reactions and the development of avalanche-like
electromagnetic cascades \cite{Bell:2008zzb,Fedotov:2010ja}.
It should also have some observable manifestations.

The most striking difference between the nonequilibrium spectrum of produced
particles and the equilibrium one is a significant non-monotonicity of
$f(\mathbf{p})$.
This deformation is much more pronounced in the mid-state which is important for
elementary reactions and cascades.
This feature has been noticed in the works \cite{Blaschke:2005hs,Filatov:2006}
for the case of a periodic field, and was later studied more in detail for a
Gaussian laser pulse in the work \cite{Hebenstreit:2009km}.
The representative oscillations of the momentum distribution with a scale set
by the laser frequency received an explanation on the mathematical basis for
the Stokes phenomenon in the theory of differential equations
\cite{Dumlu:2010ua}.
Other manifestations of such effects were studied in
\cite{Labun:2011xt,Hebenstreit:2011wk}.

It is important to note that these remarkable features of the momentum
distribution function are not captured by WKB-like approaches
\cite{Brezin:1970xf,Popov:2001}.
On the other hand, the only known exact solution for a homogeneous field in
the form of a Sauter bell generates a smooth spectrum \cite{Nikishov:1970}.
Apparently, for the occurrence of nonmonotonic deformations of $f(\mathbf{p})$
different scales must be present in the spectrum of the external field.
The actual laser field is required in the spectrum of two different time
scales, so that the question of the form $f(\mathbf{p})$ is essential for the
experiment.

In the framework of using limitations, the following main results were obtained
by both numerical and analytical approaches:
\begin{itemize}
  \item the distribution function of the residual EPPP has a complicated
quasiperiodical character in both the longitudinal ($p_\parallel$) and the
transversal ($p_\bot$) {momentum components} resulting in a cellular structure {in the momentum space};
  \item the profile function for switching-on (and switching-off) the laser
signal has a considerable influence on the magnitude and structure of the
momentum distribution of the residual EPPP;
  \item together with the usual multiphoton mechanism of the EPPP excitations
the photon cluster mechanism of absorption of the laser field energy plays an
essential role due to the relativistic nature of this system.
   \end{itemize}

In Sect.~\ref{sect:low} we explain the problem and give an analytic solution of
the KE in the domain $\gamma \gg 1$ and $E_0 \ll E_c$.
Such analytical estimations are important due to the complications encountered
in the numerical calculations.
In Sect.~\ref{sect:num} we present some numerical solutions of the KE in this
domain. The basic results are summarized in Sect.~\ref{sect:sum}.

\section{Low density limit in the multiphoton domain 
($\gamma \gg 1, E_0 \ll E_c$)
\label{sect:low}}
For the description of the EPPP vacuum creation we use the exact
nonperturbative KE obtained in the work \cite{Schmidt:1998vi} for an arbitrary
time dependent and spatially homogeneous electric field with linear
polarization for which we use the Hamiltonian gauge $A^\mu(t)=(0,0,0,A(t))$
and the field strength $E(t)=-\dot{A}(t)$,
\begin{equation}\label{ke}
\dot f(\mathbf{p},t) = \frac{1}{2} \lambda(\mathbf{p},t)\int\limits^t_{t_0}
dt^\prime \lambda(\mathbf{p},t^\prime)[1-2f(\mathbf{p},t^\prime)]
\cos\theta(t,t^\prime),
\end{equation}
where
\begin{equation}
 \lambda(\mathbf{p},t) = e E(t)\varepsilon_{\bot}/\omega^{2}(\mathbf{p},t)
\label{3}\end{equation}
is the amplitude of the EPPP excitations,
$\omega(\mathbf{p} ,t) = \sqrt{\varepsilon^2_{\bot}(\mathbf{p}) +
(p_{\parallel}-eA(t))^2}$ with the transverse energy
$\varepsilon_\bot = (m^2 + p^2_\bot)^{1/2}$  and the high frequency phase is
\begin{equation}
\theta(t,t^\prime) = 2 \int^t_{t^\prime} d\tau \omega(\mathbf{p} ,\tau) ~.
\label{phase}
\end{equation}
The distribution function in the quasiparticle representation 
$f(\mathbf{p} ,t) = <in\mid a^{+}(\mathbf{p} ,t)a(\mathbf{p} ,t)\mid in>$ 
is defined with the in-vacuum state $\mid in>$.
For the generalization of the KE (\ref{ke}) to arbitrary electric field
polarization see \cite{Filatov:2006,Pervushin:2006vh,Filatov:2009xd}.

In the low-density limit $f\ll 1$, the KE (\ref{ke}) leads to the formal
solution
\begin{equation}
f_{\rm low}(\mathbf{p},t)=\frac{1}{4} \biggl| \int\limits^t_{-\infty}dt^\prime
\lambda(\mathbf{p} ,t^\prime) {\rm e}^{i\theta(t,t^\prime)}\biggr|^{2}~.
\label{ld1}
\end{equation}
Here, the external field is switched on in the infinite past,
$t_0 \to -\infty$.
The approximation (\ref{ld1}) implies that the external field is rather weak, $E \ll E_c$.
Below we will restrict ourselves to the analysis of Eq.~(\ref{ld1}) in the
limit $t \to \infty$, which defines the momentum distribution of the residual
EPPP,
\begin{equation}
f_{\rm out}(\mathbf{p}) = \lim_{t \to \infty} f_{\rm low}(\mathbf{p} ,t)
= \frac{1}{4} \biggl| \int\limits_{-\infty}^{\infty}dt \lambda(\mathbf{p} ,t)
{\rm e}^{i\theta(t)}\biggr|^{2}~.
\label{ld2}
\end{equation}
Here the representation of the phase (\ref{phase}) via antiderivatives has been
used, $\theta(t,t^\prime) = \theta(t) - \theta(t^\prime)$.

Hereafter we will consider the domain of the multiphoton mechanism of EPPP
vacuum creation, where the adiabaticity parameter is large \cite{Popov:2001},
$\gamma \gg 1$.
A perturbation theory w.r.t. the small parameter $1/\gamma \ll 1$ can be
constructed here.

As a first step let us consider the function $f_{\rm out}(\mathbf{p})$
(\ref{ld2}) in the leading approximation, when
\begin{eqnarray}
\omega(\mathbf{p} ,t) &\to& \omega_0(\mathbf{p})= \sqrt{\mathbf{p}^2 + m^2};
\nonumber \\
\lambda(\mathbf{p} ,t) &\to& \lambda_0(\mathbf{p}) E(t), \qquad
\lambda_0(\mathbf{p}) = e \varepsilon_{\bot}/ \omega_0^2(\mathbf{p})~,
\label{6}
\end{eqnarray}
and $\theta(t) = 2 \omega_0 \cdot t$.
This means that the contribution of the high frequency harmonics of the
external field is omitted in Eq. (\ref{ld2}).
It leads to the simple result
\begin{equation}
f_{\rm out}(\mathbf{p}) = \frac{1}{4} \lambda_0^2 \biggl
| E(\omega = 2 \omega_0 ) \biggr|^{2}~,
\label{2w}
\end{equation}
where $E(\omega)$ is the Fourier transform of the field strength $E(t)$.
In this approximation the vacuum EPPP production takes place when for the
frequency of the one photon $e^-e^+$ pair creation process $\omega_{1\gamma}$
holds $\omega_{1\gamma} = 2 \omega_0$.
This mechanism is exclusive here.
Its intensity is regularized by the presence of the frequency
$\omega_{1\gamma}$ in the spectrum of an external field.

In order to facilitate EPPP creation, one can take into account the
multiphoton mechanisms of EPPP production.
These processes are described by the high frequency multiplier in
Eq.~(\ref{ld2}).
Let us use the non-perturbative method of photon counting assuming that the
electric field is periodical, $A(t)=A(t+2\pi /\nu)$, where $\nu =2\pi/T$ is
the angular frequency and $A(t)= A(-t)$, in order to provide the property of
being an even function of the quasiparticle energy,
$\omega(\mathbf{p} ,t) = \omega(\mathbf{p} ,-t)$.
The corresponding Fourier transform
\begin{equation}\label{furw}
\omega(\mathbf{p} ,t) = \sum\limits_{n=1} \Omega_n \cos{n\nu t}
\end{equation}
leads to the decomposition of the phase in Eq. (\ref{ld2})
\begin{equation}\label{furt}
\theta(t) = 2\Omega_0 t + \sum\limits_{n=1} a_n \sin{n\nu t},
\end{equation}
where $a_n = 2\Omega_n/\nu n$.
In the case of a harmonic external field $\Theta_0$ is the renormalized
frequency \cite{Brezin:1970xf}.
Let us employ now the leading approximation
$\lambda(\mathbf{p} ,t) \to \lambda_0(\mathbf{p}) E(t)$ (\ref{6}) and a
non-perturbative decomposition based on the well known formula
\begin{equation}\label{bess}
    \exp{(i a \sin{\phi})} =
\sum\limits_{n= -\infty}^{\infty} J_n (a) \mathrm{e}^{i n \phi} ~,
\end{equation}
where $J_n (a)$ is the Bessel function.

Let us now consider the integral in Eq.~(\ref{ld2}) and perform the
substitutions (\ref{furw})-(\ref{bess}), leading to
\begin{eqnarray}
J &=& \int\limits_{-\infty}^{\infty} dt E(t)\, {\rm e}^{i\theta(t)} \nonumber\\
&=& \int\limits_{-\infty}^{\infty} dt E(t)\, {\rm e}^{2i\Omega_0 t}
\biggl\{ J_0 (2 \Omega_0)
+  \prod\limits_{n=1} \sum_{k_n =1} J_{k_n} (a_n)\, {\rm e}^{i(k_n n) \nu t}
+ (k_n \to -k_n) \biggr\}~.
\label{11}
\end{eqnarray}
The first term with $J_0 (2 \Omega_0)$ describes the direct vacuum excitations
at the frequency $\omega = 2\Omega_0$ and corresponds to Eq.~(\ref{2w}).
If we are interested in excitations at lower frequencies, this contribution
can be omitted.

A new feature of the multiphoton processes here is the cluster character of
the energy absorption from the photon reservoir of the external field.
The photon cluster of n$^{\rm th}$ order consists of $n$ identical photons with
the carrier frequency $\nu$ and the total energy $n \nu$.
As it can be seen from Eq.~(\ref{11}), it is possible that $k$ of the
$n$-photon clusters are absorbed simultaneously.
The ordinary multiphoton process corresponds to the simplest "cluster" of the
order $k=1$.
The appearance of photon clusters in the multiphoton processes in
Eq.~(\ref{11}) is a consequence of the nonlinear field dependence of the
quasiparticle energy $\omega(\mathbf{p} ,t)$.
The probability for the generation of an $n$-photon cluster is defined by the
amplitude $a_n$ in the decomposition (\ref{furt}) (the argument of the Bessel
function $J_k(a_n)$ in Eq. (\ref{11})) while the probability of the
simultaneous $k$-cluster absorption corresponds to the Bessel function of the
order $k$.

We rewrite the integral (\ref{11}) in the following approximation:
\begin{align}
J(n_{\rm max}) = \int\limits_{-\infty}^{\infty} dt
E(t)\, {\rm e}^{2i\Omega_0 t} \prod\limits_{n=1} \sum\limits_{k_n =1}
\biggl\{  J_{k_n} (a_n)\, \exp{ \biggl( i \sum\limits_{n =1}^{n_{max}}
(n k_n) \nu t \biggr) } + (k_n \to -k_n) \biggr\} \label{12}~,
\end{align}
where $n_{\rm max}$ is the maximal photon number in the cluster,
$J= J(n_{\rm max}\to \infty)$.
 Let us perform here the resummation procedure.
In Eq.~(\ref{12}) we conserve the index $k_n$ for notation of the $k_n$
identical photon clusters where each cluster contains $n$ photons.
Then the total photon number in this photon set will be equal to $N_k=n k_n$.
Let us replace now the summation over $n$ by a summation over $N_k$.
The integral (\ref{12}) can then be rewritten as
\begin{align}
\label{13}
    J = 2 \pi \sum\limits_{k=1}
E(\omega = 2\Omega_0 - \nu \sum\limits_{N_k  =1}^{N_{\rm max}} [N_k])
\prod\limits_{N_k =1} J_k (a_{N_k/k})~.
\end{align}
Here $N_k$ denotes an arbitrary integer multiple of $k$ and
$N_{\rm max} \sim n_{\rm max}$ in Eq.~(\ref{12}).

Substituting Eq.~(\ref{13}) into Eq.~(\ref{ld2}), we obtain the generalization
of Eq.~(\ref{2w})
\begin{align}
\label{14}
f_{\rm out}(\mathbf{p})=\pi^2 \lambda_0^2 \biggl|
\sum\limits_{k=1} E(\omega = 2\Omega_0 - \nu \sum\limits_{N_k=1}^{N_{\rm max}}
[N_k]) \prod\limits_{N_k =1} J_k (a_{N_k/k})
\biggr|^{2}~.
\end{align}
The summation is developed here by the order of the photon cluster.
Below we will restrict the sum to the usual multiphoton processes ($k=1$).

Thus, the multiphoton mechanism leads to a lowering of the EPPP excitation
frequency.
The maximal effect is reached on the carrier frequency $\nu$ of the external
field, i.e. for the condition
\begin{equation}\label{15}
    2\Omega_0 -\nu \sum\limits_{N_k  =1}^{N_{\rm max}} [N_k] = \nu.
\end{equation}

Let $\nu \ll m$. Then the hypothesis of the phase randomization can be used,
\begin{equation}
\label{16}
    \prec E(N \nu) E^*(N'\nu)\succ_{\Delta \omega}=|E(N \nu)|^2 \delta_{NN'}~,
\end{equation}
where $N, N'$ are integers and $\prec \dots \succ_{\Delta \omega}$ denotes the
averaging procedure over a small frequency interval $\Delta \omega$.
The approximation (\ref{16}) applied to Eq.~(\ref{14}) leads to the result
\begin{align}
\label{17}
 f_{\rm out}(\mathbf{p}) =
\pi^2 \lambda_0^2 \sum\limits_{k=1} \biggl|
E(2\Omega_0 - \nu \sum\limits_{N_k  =1}^{N_{\rm max}} [N_k])
\biggr|^{2} \prod\limits_{N_k =1} J^2_k (a_{N_k/k}) ~.
\end{align}

In the case of the monochromatic external field Eq.~(\ref{17}) exhibits the
accumulation effect.
Using the formula $\delta^2(\omega) = (T/2\pi)\cdot  \delta(\omega)$ for
$T \gg 2\pi/\nu$, one can obtain from Eq.~(\ref{17}) the EPPP production rate
\begin{align}
\label{18}
    I_{\rm out}(\mathbf{p}) =
\frac{1}{2} \pi E_0^2 \lambda_0^2\, \sum\limits_{k=1} \prod\limits_{N_k =1}
J^2_k (a_{N_k/k})\, \delta \bigl(2\Omega_0
- \nu \sum\limits_{N_k  =1}^{N_{\rm max}} [N_k] -\nu \bigr)~.
\end{align}

Then the condition (\ref{15}) defines the photon number required for the
breakdown of the energy gap with the renormalized mass
$m_{\rm ren} = \Omega_0(\mathbf{p}=0)$.
Since the renormalized frequency $\Omega_0(\mathbf{p})$ is a monotonously
increasing function of $p=|\mathbf{p}|$ with
min $\Omega_0(\mathbf{p}) = m_{\rm ren}$,
this photon number will be minimal for the particles created at rest,
$N_0 = N_{\rm max}(p=0)$.
In order to investigate the momentum distribution of the EPPP it is
necessary to consider a larger photon number,
$N_{\rm max} > N_0$ and $N_{\rm max} \to \infty$ for $p \to \infty$.

The simplest situation corresponds to the usual multiphoton processes when the
photon cluster mechanism is a minor, $k=1$.
The Eq.~(\ref{18}) in this case has the form
\begin{align}
\label{19}
    I_{\rm out}(\mathbf{p})=\frac{1}{2}\pi E_0^2~\lambda_0^2~\delta~\bigl[
2\Omega_0(\mathbf{p})-\nu (N_{\rm max}+1)
\bigr]\prod\limits_{n = N_0}^{N_{\rm max}} J^2_1 (a_{n}) ~.
\end{align}
The set of the Bessel function arguments $a_n$ corresponds here to the spectral
composition of the multiphoton process in Eqs.~(\ref{furw}) and (\ref{furt}).

Already on the qualitative level it can be seen from Eqs.~(\ref{18}) and (\ref{19}) that the momentum distribution of the residual EPPP has a cellular structure in momentum space stipulated by the presence of the Bessel function.
It leads to some quasi-periodic behavior of $f_{\rm out}(\mathbf{p})$
w.r.t. the components $p_\parallel$ and $p_\bot$.

Eqs.~(\ref{18}) and (\ref{19}) allow to perform various estimations of EPPP
characteristics (particle number density, momentum distributions) for different parameters of the laser radiation in the domain $\gamma \gg 1$.

As an example we find the estimate for the total intensity of EPPP production
\begin{equation}
\label{20}
I = 4 \int\limits \frac{d^3 p}{(2 \pi)^3}  I_{\rm out}(\mathbf{p})
\end{equation}
for $\nu = m$ and $E_0=0.1~E_c$.
It corresponds to the two-photon mechanism of EPPP creation.
From Eq.~(\ref{19}) it follows then that the rate of EPPP production in the
volume $\lambda^3$ per period of the field action ($E_0/E_c = \xi$) is
\begin{equation}
\label{21}
    I = \frac{1}{2 \pi} \xi^2 J_1^2(\xi^2/4) m^4
\approx 10^{-9}/(\lambda^3 T) ~.
\end{equation}

\begin{figure}[t]
\begin{minipage}[t]{0.47\textwidth}
\includegraphics[width=0.99\textwidth]{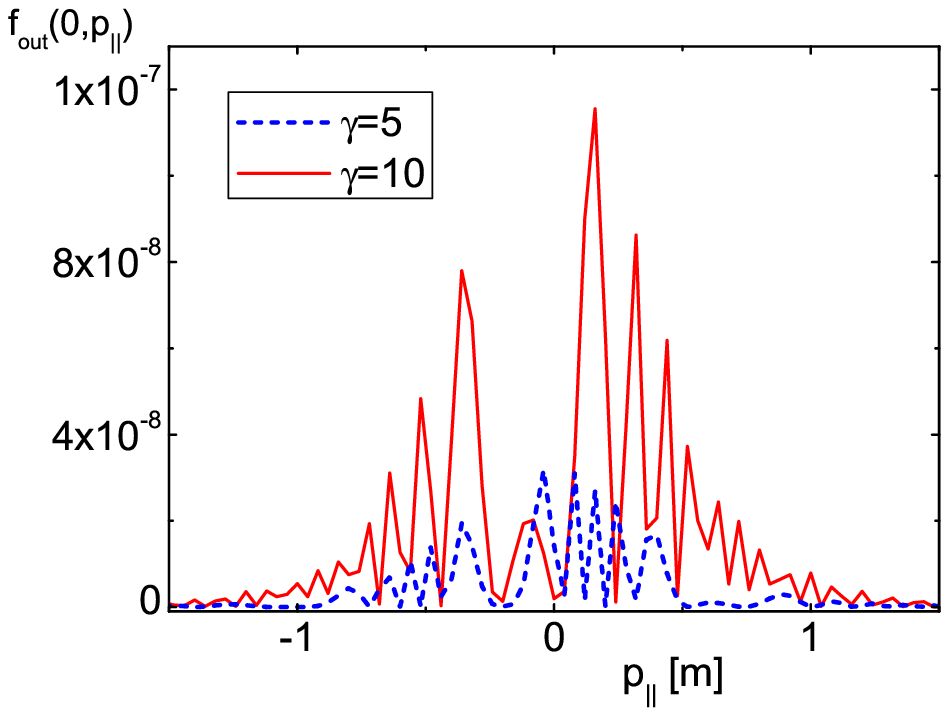}
\caption{Section $p_\bot =0$ of $f_{\rm out}(\mathbf{p})$ for the pulse shape
(\ref{radio}) for two values of $\gamma$.
\label{fig1}}
\end{minipage}
\hspace{\fill}
\begin{minipage}[t]{0.47\textwidth}
\includegraphics[width=0.99\textwidth]{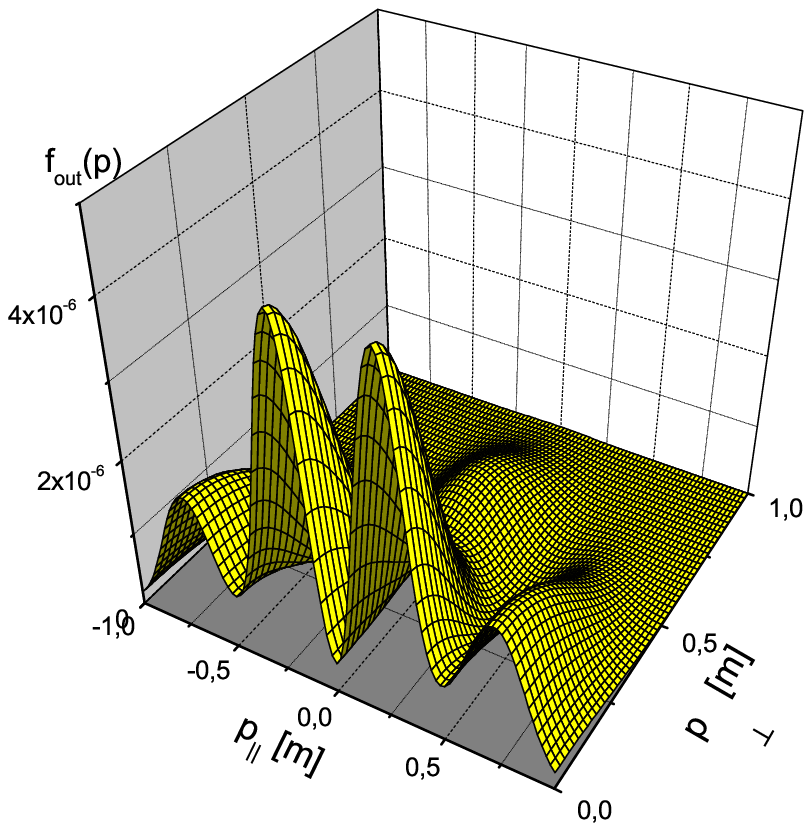}
\caption{The shape of $f_{\rm out}(\mathbf{p})$ for a Gaussian pulse
(\ref{gauss}) with $\gamma=10$, $\sigma=5$ and $E_0=0.1~E_c$.
\label{fig2}}
\end{minipage}
\end{figure}

\section{Numerical calculations \label{sect:num}}

\begin{figure}[h]
\begin{minipage}[t]{0.47\textwidth}
\includegraphics[width=0.99\textwidth]{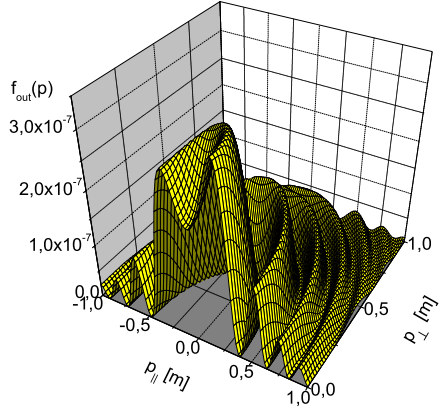}
\end{minipage}\hspace{\fill}
\begin{minipage}[t]{0.47\textwidth}
\includegraphics[width=0.99\textwidth]{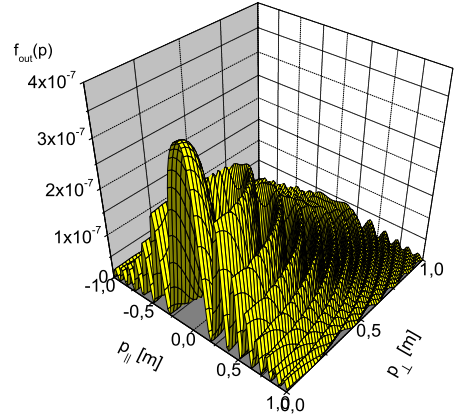}
\end{minipage}
\caption{Increasing complexity of $f_{\rm out}(\mathbf{p})$ in the field
(\ref{radio}) with increasing $N$ at $E_0=0.01~E_c$ and  $\gamma=24$.
Left panel: N=1; right panel: N=2.
\label{fig34}}
\end{figure}

We explore field pulses with two types of shape: the radio pulse
($N$ periods of the harmonic field)
\begin{equation}
\label{radio}
E(t) = E_0 \sin{\nu t}, \quad 0 \le t \le N\, 2\pi/\nu, \quad N-\mbox{integer},
\end{equation}
and the Gaussian pulse
\begin{equation}
E(t) = E_0 \exp{(-t^2/2\tau^2)} \cos{\nu t}, \quad \nu \tau =\sigma,\quad
-10\tau \le t \le 10\tau~.
\label{gauss}
\end{equation}

\begin{figure}[t]
\begin{minipage}[t]{0.47\textwidth}
\includegraphics[width=0.8\textwidth]{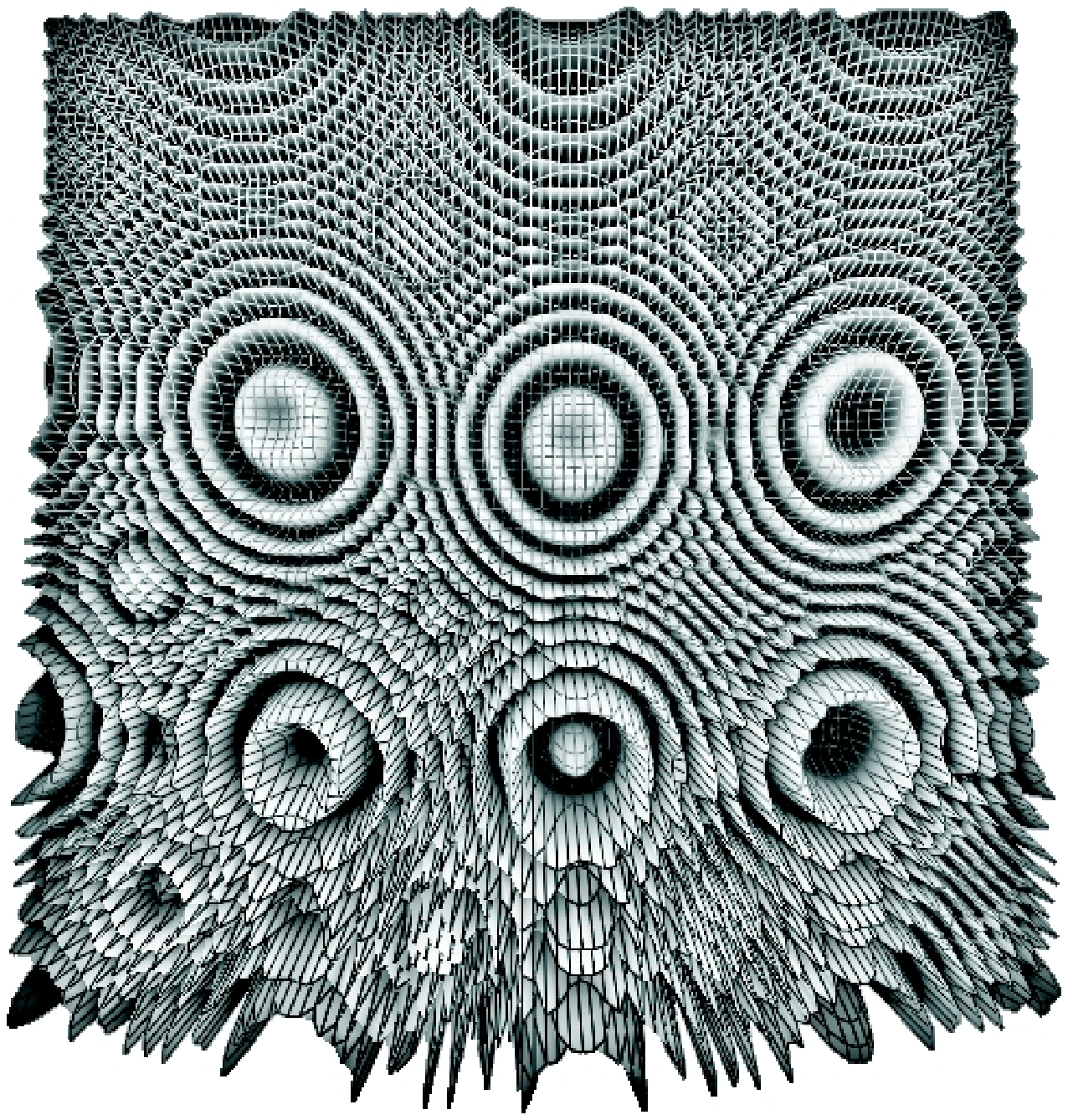}
\end{minipage}\hspace{\fill}
\begin{minipage}[t]{0.47\textwidth}
\includegraphics[width=0.8\textwidth]{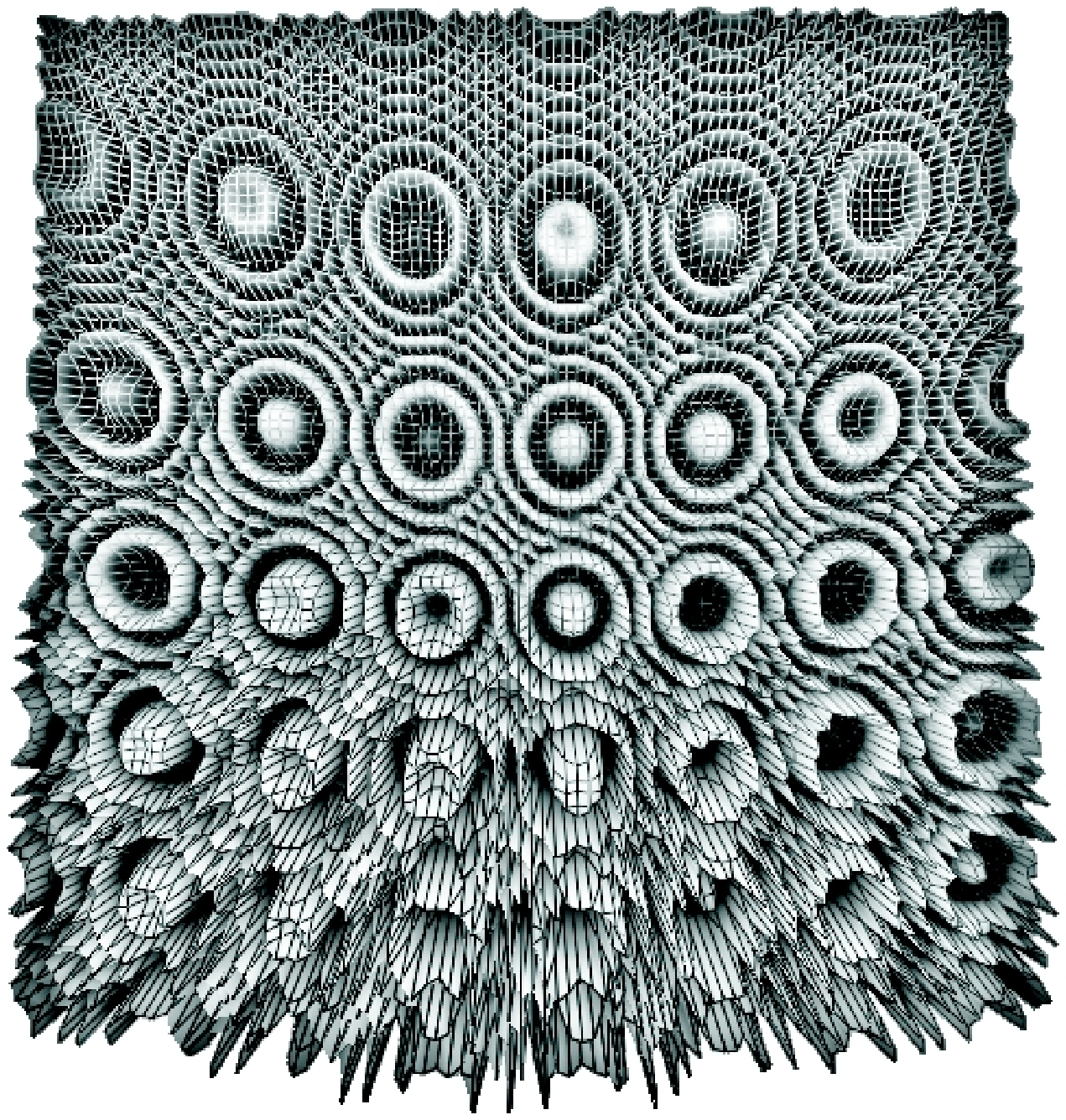}
\end{minipage}
\caption{The complex cellular structure of the shape of
$f_{\rm out}(\mathbf{p})$  at large N in a survey top view.
Left panel: left N=50; right panel: N=100.
\label{fig56}}
\end{figure}

Figures \ref{fig1}, \ref{fig34}, \ref{fig56} show the complicated behaviour of
the momentum distribution $f_{\rm out}(\mathbf{p})$ for the periodical field
(\ref{radio}) in dependence on the field parameters.
This distribution turns out to be asymmetric w.r.t. the interchange
$p_{||} \to - p_{||}$ (Fig. \ref{fig1}).
Increasing the number $N$ of periods in the pulse (\ref{radio}) leads to a
complication of the momentum distribution (Fig.~\ref{fig34}) and to the
generation of a cellular structure for large $N \gg 1$ (Fig.~\ref{fig56}).
These  details are lost when using the WKB methods 
\cite{Brezin:1970xf,Popov:2001}.

\begin{figure}[b]
\begin{minipage}[t]{0.47\textwidth}
\includegraphics[width=0.99\textwidth]{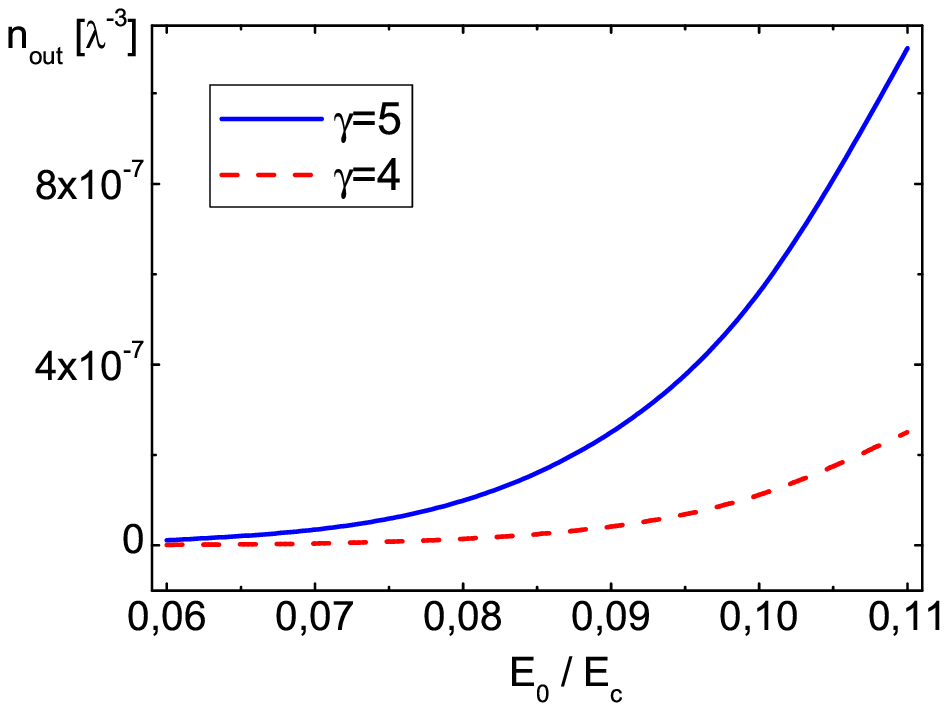}
\caption{Dependence of $n_{\rm out}$ on the field strength for a Gaussian pulse
(\ref{gauss}) for two values of $\gamma$.
\label{fig7}}
\end{minipage}\hspace{\fill}
\begin{minipage}[t]{0.47\textwidth}
\includegraphics[width=0.99\textwidth,height=0.77\textwidth]{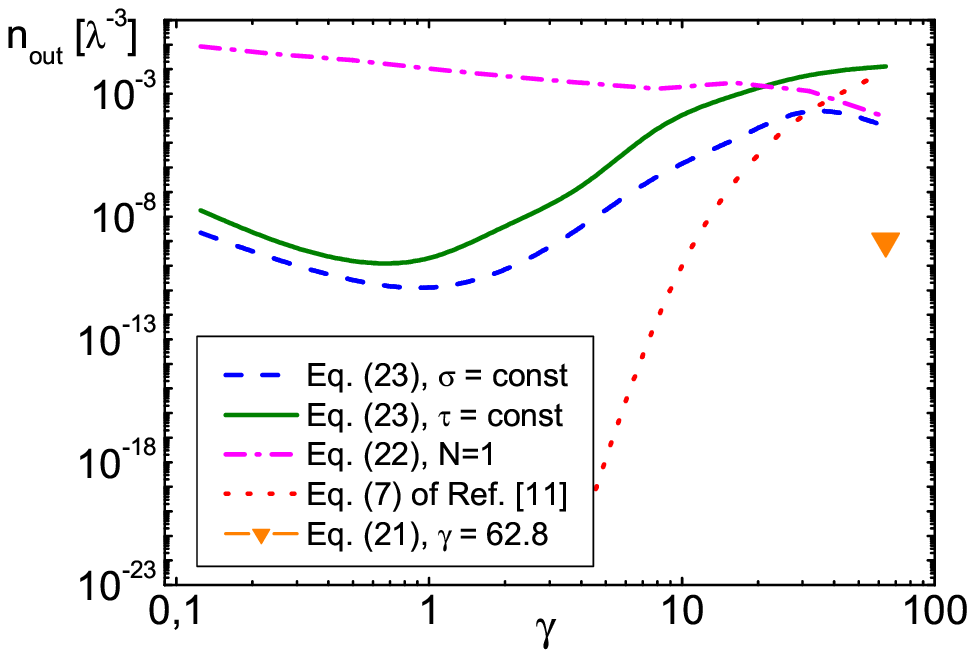}
\caption{Comparison of $n_{\rm out}$ for different pulse shapes and different 
theoretical approaches.
\label{fig8}}
\end{minipage}
\end{figure}

Analogous alterations can be observed in the case of the Gaussian pulse
(\ref{gauss}).
For example, Fig.~\ref{fig2} shows the distribution function for $\sigma = 5$
and $\gamma = 10$.
Fig.~\ref{fig7} illustrates the strong dependence of $n_{\rm out}$ on the
field strength.
A comparison of the EPPP number density for the pulses (\ref{radio}) and
(\ref{gauss}) shows that the periodical field can produce a significantly
higher output of observable pairs.
The obtained estimates for $n_{\rm out}$ are notably different also from the
well-known results of works based on the WKB approach.

\clearpage
\section{Conclusions \label{sect:sum}}

In the present work we have done first steps in the investigation of the
residual EPPP based on a kinetic theory foundation for strong nonperturbative
QED.
We restricted ourselves here to the multiphoton domain $\gamma \gg 1$ and the
low-density approximation.
We have shown that the momentum distribution of the residual EPPP is a strong
nonequilibrium one and has a complicated cellular structure in the momentum
space depending on the characteristics of the laser radiation.
As a result of the numerical analysis it was observed that the residual EPPP
density depends strongly on the pulse shape for switching on
(and switching off) the laser field.
Finally, a theoretical approach was developed for estimating different
characteristics of the residual EPPP.
This method allows to take into account both the usual multiphoton processes
of EPPP excitation and also the new photon cluster mechanism.

\begin{acknowledgement}
S.A.S. acknowledges support by Deutsche Forschungsgemeinschaft (DFG)
under Project number TO 169/16-1.
D.B. is grateful for hospitality and financial support during his stay at the
University of Bielefeld where this work was finalized.
\end{acknowledgement}


\begin{thebibliography}{99}

\bibitem{Blaschke:2005hs}
D.~B.~Blaschke, A.~V.~Prozorkevich, C.~D.~Roberts, S.~M.~Schmidt and
S.~A.~Smolyansky, 
Phys.\ Rev.\ Lett.\  {\bf 96}, 140402 (2006).

\bibitem{Blaschke:2008wf}
  D.~B.~Blaschke, A.~V.~Prozorkevich, G.~R\"opke, C.~D.~Roberts,
S.~M.~Schmidt, D.~S.~Shkirmanov and S.~A.~Smolyansky,
  Eur.\ Phys.\ J.\ D {\bf 55}, 341 (2009).

\bibitem{Gregori:2010uf}
G.~Gregori, D.~B.~Blaschke, P.~P.~Rajeev, H.~Chen, R.~J.~Clarke, T.~Huffman,
C.~D.~Murphy, A.~V.~Prozorkevich, C.~D.~Roberts, G.~R\"opke,
S.~M.~Schmidt, S.~A.~Smolyansky, S.~Wilks, R. Bingham,
  High Energy Dens.\ Phys.\  {\bf 6}, 166 (2010).

%
\bibitem{Ringwald:2001ib}
  A.~Ringwald,
  Phys.\ Lett.\ B {\bf 510}, 107 (2001).

\bibitem{desy}
M.~Altarelli, et al. (Eds.),
{\it XFEL, The European X-ray Free-Electron Laser},
Technical Design Report, DESY 2006-097, Hamburg (2006).

\bibitem{Bulanov:2003zz}
  S.~V.~Bulanov, T.~Esirkepov and T.~Tajima,
  Phys.\ Rev.\ Lett.\  {\bf 91}, 085001 (2003)
  [Erratum-ibid.\  {\bf 92}, 159901 (2004)].

\bibitem{Rhodes}
C.~K.~Rhodes, J.~Boguta, A.~B.~Borisov, S.~F.~Khan,
J.~W.~Longworth, J.~C.~McCorkindale, S.~Poopalasingam,
E.~Racz, J.~Zhao //
Reaching the Schwinger Limit with X-Rays //
Proc. of International Conference on Physics in Intense Fields (PIF 2010),
24 - 26 November, 2010, KEK, Tsukuba, Japan.

\bibitem{Popov:2001}
V.~S.~Popov, Sov. Phys. JETP, {\bf 120}, 315 (2001).

\bibitem{Brezin:1970xf}
  E.~Brezin and C.~Itzykson,
  Phys.\ Rev.\  D {\bf 2}, 1191 (1970).

\bibitem{Schmidt:1998vi}
S.~M.~Schmidt, D.~Blaschke, G.~R\"opke, S.~A.~Smolyansky, A.~V.~Prozorkevich
and V.~D.~Toneev,
Int.\ J.\ Mod.\ Phys.\ E \textbf{7}, 709 (1998).

\bibitem{Schmidt:1998zh}
  S.~M.~Schmidt, D.~Blaschke, G.~R\"opke, A.~V.~Prozorkevich,
S.~A.~Smolyansky and V.~D.~Toneev,
  Phys.\ Rev.\ D {\bf 59}, 094005 (1999).

\bibitem{Blaschke:2011is}
  D.~B.~Blaschke, V.~V.~Dmitriev, G.~R\"opke,  S.~A.~Smolyansky,
  Phys.\ Rev.\  {\bf D84}, 085028 (2011).



\bibitem{Bell:2008zzb}
  A.~R.~Bell and J.~G.~Kirk,
  Phys.\ Rev.\ Lett.\  {\bf 101}, 200403 (2008).

\bibitem{Fedotov:2010ja}
  A.~M.~Fedotov, N.~B.~Narozhny, G.~Mourou and G.~Korn,
  Phys.\ Rev.\ Lett.\  {\bf 105}, 080402 (2010).

\bibitem{Filatov:2006}
A.~V.~Filatov, A.~V.~Prozorkevich, and S.~A.~Smolyansky,
Proc. of SPIE, {\bf 6165}, 616509 (2006).

\bibitem{Hebenstreit:2009km}
  F.~Hebenstreit, R.~Alkofer, G.~V.~Dunne and H.~Gies,
  Phys.\ Rev.\ Lett.\  {\bf 102}, 150404 (2009).

\bibitem{Dumlu:2010ua}
  C.~K.~Dumlu and G.~V.~Dunne,
  Phys.\ Rev.\ Lett.\  {\bf 104}, 250402 (2010).

  \bibitem{Labun:2011xt}
  L.~Labun and J.~Rafelski,
  Phys.\ Rev.\ D {\bf 84}, 033003 (2011).

\bibitem{Hebenstreit:2011wk}
  F.~Hebenstreit, R.~Alkofer and H.~Gies,
  Phys.\ Rev.\ Lett.\  {\bf 107}, 180403 (2011).

 \bibitem{Nikishov:1970}
  N.~B.~Narozhnyi and A.~I.~Nikishov, Yad. Fiz. {\bf 11} (1970) 1072
  [Sov. J. Nucl. Phys. {\bf 11} (1970) 596].

\bibitem{Pervushin:2006vh}
  V.~N.~Pervushin, V.~V.~Skokov,
  Acta Phys.\ Polon.\  {\bf B37}, 2587 (2006).

\bibitem{Filatov:2009xd}
  A.~V.~Filatov, S.~A.~Smolyansky, and A.~V.~Tarakanov,
 Proc. of the XX International Baldin Seminar on High Energy Physics Problems "Relativistic Nuclear Physics and Quantum Chromodynamics",
Dubna, Sept. 29 - Oct. 04, 2008, 202 (2008).



\end{thebibliography}
\end{document}